\documentclass{aa}

\usepackage[varg]{txfonts}

\usepackage{hyperref}

\usepackage{epic}
\usepackage{eepic}
\usepackage{graphicx}        
\usepackage{amssymb}         
\usepackage[usenames,dvipsnames]{color}          

\bibpunct{(}{)}{;}{a}{}{,} 

\usepackage{epstopdf}
\usepackage{amssymb,amsmath}

\newcommand{\rd}{\mathrm{d}}

\newcommand{\pderiv}[2]{\frac{\partial#1}{\partial#2}}

\newcommand{\deriv}[2]{\frac{\rd#1}{\rd#2}}

\newcommand{\g}{\mathbf{g}}
\newcommand{\boldk}{\mathbf{k}}

\newcommand{\x}{\mathbf{x}}

\newcommand{\vdot}{{\boldsymbol{\cdot}}}

\newcommand{\thth}{\hspace{1.5pt}}

\newcommand{\kpar}{k_{\scriptscriptstyle\parallel}}

\newcommand{\bv}{Brunt-V\"ais\"al\"a}

\newcommand{\calD}{{\mathcal{D}}}

\allowdisplaybreaks[1]

\begin{document}


\title{Probing sunspots with two-skip time-distance helioseismology}

\author{Thomas L. Duvall Jr.\inst{\ref{inst1}}
\and Paul S. Cally\inst{\ref{inst2}}
\and Damien Przybylski\inst{\ref{inst2},\ref{inst1}}
\and Kaori Nagashima\inst{\ref{inst1}}
\and Laurent Gizon\inst{\ref{inst1},\ref{inst3}}
}

\institute{Max-Planck-Institut f\"ur Sonnensystemforschung, Justus-von-Liebig-Weg 3, 37077 G\"ottingen, Germany\label{inst1}
\and
School of Mathematical Sciences and Monash Centre for Astrophysics, Monash University, Clayton, Victoria 3800, Australia\label{inst2}
\and
Institute f\"ur Astrophysik, Georg-August-Universt\"at G\"ottingen, Friedrich-Hund-Platz 1, 37077 G\"ottingen, Germany\label{inst3}
}

\authorrunning{Duvall, Cally, Przybylski, Nagashima, \& Gizon}

\titlerunning{Probing Sunspots with Two-Skip Helioseismology}

\date{Received <date> / Accepted <date>}

\abstract
{ Previous helioseismology of sunspots has been sensitive to both the
structural and magnetic aspects of sunspot structure.}
{ We aim to develop a technique that is insensitive to the magnetic component 
so the two aspects can be more readily separated.}
{ We study waves reflected almost vertically from the underside of a sunspot.
Time-distance helioseismology was used to measure travel times
for the waves.  Ray theory and a detailed sunspot model were used to 
calculate travel times for comparison.}
{ It is shown that these large distance waves are insensitive to the
magnetic field in the sunspot.  The largest travel time differences for any
solar phenomena are observed. } 
{ With sufficient modeling effort, these should
lead to better understanding of sunspot structure.}


\maketitle


\section{Introduction}

\citet{Schunker13} have shown that relatively shallow, horizontally
propagating, f and p modes have sensitivity to both the magnetic and
thermal structure of a sunspot.  They found that travel-time measurements
can constrain the height of the Wilson depression to a precision of ~50 km.
\citet{Lindsey10} showed that rays approaching the sunspot
almost vertically from below are rather insensitive to the magnetic field.
Rays approaching vertically from below may only be sensitive to the thermal
structure, or the Wilson depression.  To develop a helioseismic method
to reliably measure the Wilson depression would be a significant advance,
considering all the controversy surrounding the helioseismic sunspot
measurements \citep{Gizon09,Moradi10}.
\citet{Lindsey10} used the signal in the sunspot to
cross correlate with the ingression and egression holography signals to
get travel time perturbations.  This method has the disadvantage of using
the signal in the sunspot, which puts an additional level of uncertainty
on the results.
\citet{Chou00} computed the cross covariance between the ingression and
the egression signal
to derive travel time perturbations.  By using the Gabor wavelet to fit
the cross covariance, both a phase time and an envelope time were derived.
\citet{Chou00} found that the envelope time yields a considerably larger
signal than the phase time, much as we find in this paper.
The envelope time also has a larger error than the phase time with the
result being that the signal to noise for the phase time is larger than
for the envelope time.  However, as we see later in Fig.~\ref{Figxc},
the envelope time signal for the
umbra is not a simple multiple of the phase time signal and so there
is hopefully independent information that can be extracted.
What we are proposing here is a technique that does not
use the signal in the sunspot, and is therefore akin to the original
sunspot work with the Hankel transform which did not use signals from the
interior of the spot \citep{Braun87} and more modern techniques, such as
the one described in \citet{Cameron08}, \citet{Schunker13}, and
\citet{Liang13} which also do not
use the signal in the sunspot.

\begin{figure*}
\centering
\includegraphics[trim=0 75 0 100]{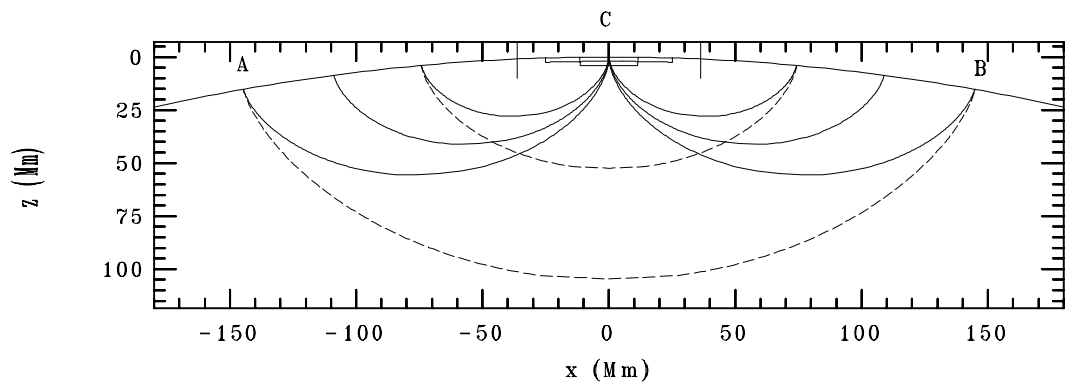}
\caption{Rays for the two-skip method in a vertical slice through the
center of the sunspot.
The umbral and penumbral locations are indicated by the (somewhat
exaggerated) pedestals near $x=0$.  The horizontal size is
derived from the ten days of data used (Nov. 14-23, 2013).  The average
umbral radius is $11.5\rm{\,Mm}$ and penumbral radius is $25.1\rm{\,Mm}$.
The solid curves
are the two-skip ray paths for the range of 1-skip distances
$\Delta=75-146\rm{\,Mm}$ used in the analysis.  The dashed curves are the
corresponding 1-skip rays.  The analysis consists of calculating temporal
cross covariances between endpoints (e.g. A and B).  For the curves drawn,
an output map point would be associated with the location half way (C) between
the endpoints, or at the center of the spot.  By moving the endpoints in
longitude and latitude, a map is constructed.  The short vertical lines
at $\pm 36\rm{\,Mm}$ indicate the size of the map shown later
in Fig.~\ref{FigIc} and Fig.~\ref{Figmap}.
}
\label{FigRay}
\end{figure*}

The new time-distance technique presented here
correlates signals from opposite sides
of the spot and uses the signal that putatively bounces halfway in
between to infer properties of the spot (Fig.~\ref{FigRay}).
That such a two-skip
signal is sensitive to the presence of the spot was first shown by
\citet{Duvall95}.  two-skip signals in sunspots were used by 
\citet{Chou09} to separate absorption, emissivity reduction and
local suppression of sources.

\begin{figure*}
\centering
\includegraphics{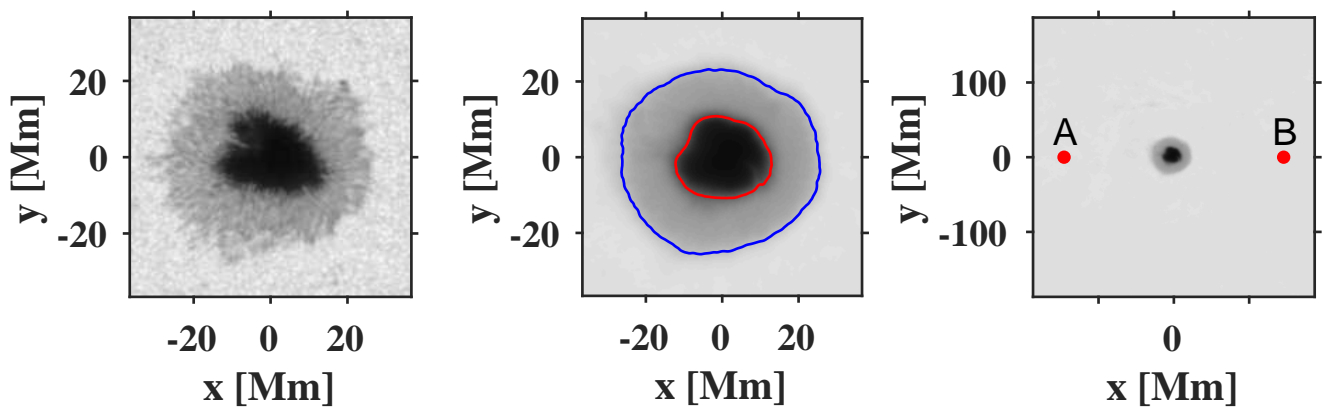}
\linebreak
\caption{
Continuum images of the sunspot in NOAA active region 11899.
The left image is a single continuum image near the central meridian passage on
Nov. 18, 2013.  The middle image is an average of the continuum images
for the ten days analyzed (Nov. 14-23, 2013).  For the left and middle
images, the horizontal size is the same as that of the eventual travel time
maps.  To identify the umbral-penumbral boundary, a contour of the
ten-day average intensity at the level of 0.4 is plotted (red).  The
penumbral-photospheric boundary is represented from the contour at
0.85 (blue).
To derive the intensities,
a fit to limb darkening is done with the sunspot excluded
and the intensity is normalized to unity
for the background photosphere.
The right image is also the ten-day average of the
continuum intensity, but showing the entire field used to derive the
travel-time maps.  The locations A, B from Fig.~\ref{FigRay} are shown.
}
\label{FigIc}
\end{figure*}

\section{Data analysis}  \label{data}
We used observations from the HMI instrument \citep{Schou12} on board the SDO
satellite.
As one of the main constraints of the present project was to use
rays that impinge on the sunspot from below in an almost vertical
direction and to not use endpoints that are in sunspots, it seemed best
to find a relatively large spot that was reasonably isolated and
did not change very much during its disk passage.  NOAA active region
11899 satisfies these requirements very nicely.  Continuum images of the
sunspot are shown in Fig.~\ref{FigIc}.  Doppler, continuum and magnetic
data from the HMI instrument \citep{Schou12} from Nov. 14-23, 2013 were
used for the sunspot analysis.  The data were broken up into ten one-day
intervals.  Each day was tracked using the program described in
\citet{Duvall13} with a sampling in longitude and latitude of $0.03\deg$,
critically sampling the HMI images at disk center.
A region covering $30\deg$ in longitude and in latitude
centered on the spot was tracked.
It was found that the computation of the cross covariances was too
time-consuming to be done with the $0.03\deg$ sampling and so the datacubes
were filtered and resampled at $0.06\deg$.  This filtering was done
by Fourier transforming the original datacubes, truncating the transforms
at half the spatial nyquist frequencies, and inverse transforming.
The central Carrington longitude, latitude,
and the rotation rate were adjusted daily to keep the spot centered.
The center of the sunspot resided in the small latitude range $5-5.1\deg$ over
the ten days.  A phase speed filter of the same form as the one applied in
\citet{Duvall13} was applied (FWHM $\Gamma=400$, units are spherical harmonic
degree $\ell$) which transmits both the
first and second skip over the range of distances used ($\Delta=12-24\deg$
of first skip distance).
This filter has central phase speed of 141 km/s.
A quiet-sun reference for the travel times was derived by doing the
same analysis on a region centered at the same latitude for the
days Nov. 8-16, 2013.  The Carrington longitude of this region at
central meridian passage is $121.8\deg$.

Cross covariance maps were computed for each of the ten days using the
program described previously \citep{Duvall03}.
This method of computing the cross covariance at opposite sides of a
circle and associating the resultant travel time with a point at depth
below the midpoint of the two locations is related to the seismic technique
of common depth point (CDP) measurements \citep{McQuillin85}.
A departure from the previous analysis is the use of eight sectors instead
of the four, or quadrants.  This enables the possibility to better study
an anisotropy of the mean signal, from which we might infer an anisotropy
of the wave speed due to the presence of a magnetic field.  For
the present study, the covariance maps have been averaged over the eight
sectors to obtain a mean signal.
The covariance maps for the different days are combined with a weighting
to remove the heliocentric angle dependence of the amplitude of the
oscillation signal.

\begin{figure*}
\centering
\includegraphics{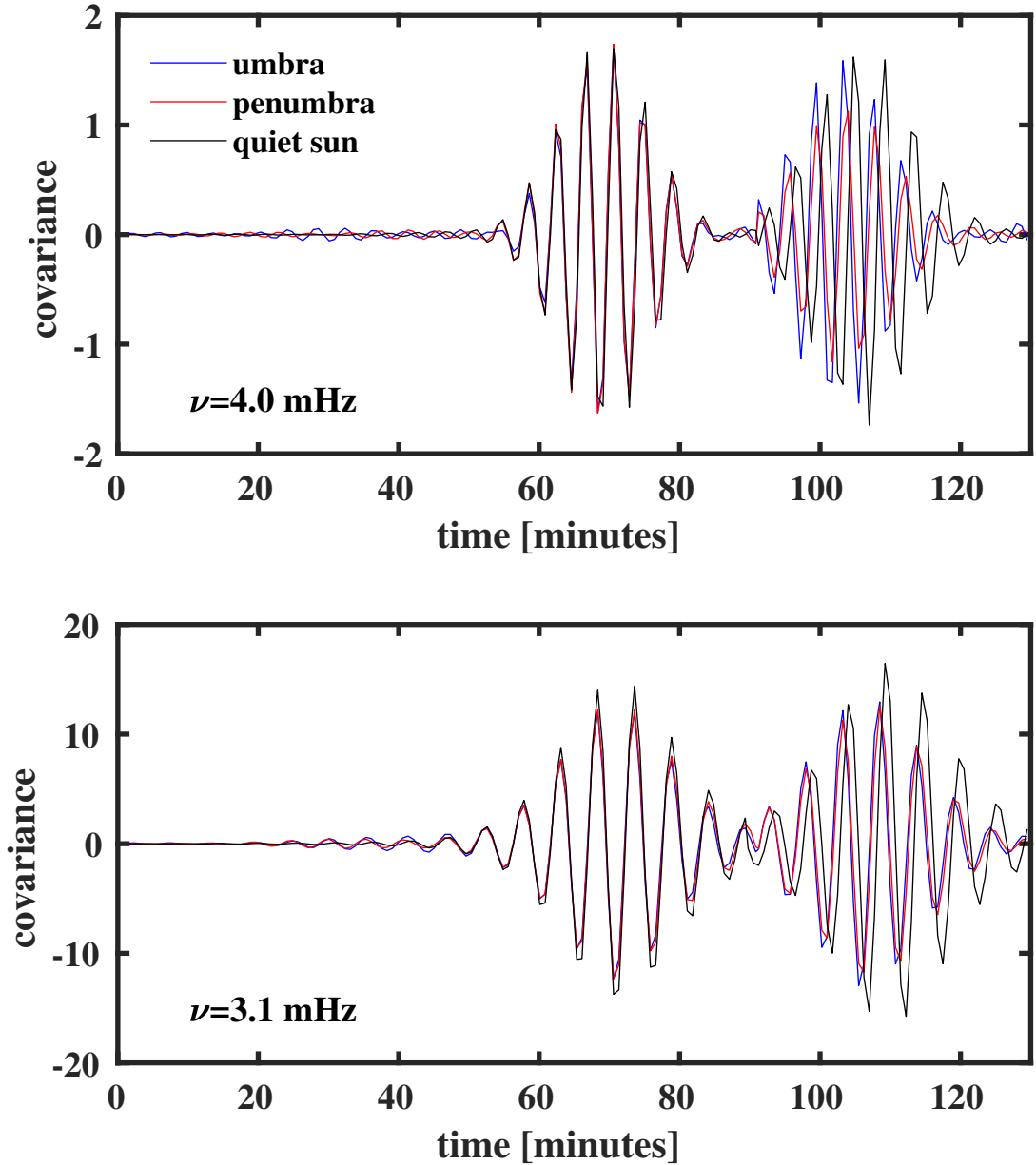}
\caption{Covariances averaged separately over the umbra, penumbra, and
the quiet Sun analysis for the full $\Delta$ range.  The top frame is
for $\nu=4.0\rm{\,mHz}$ and the bottom frame is for $\nu=3.1\rm{\,mHz}$.
The first and second skip
areas are averaged separately over $\Delta$ and then stitched together.
In both frames, the first wave packet corresponds to the first skip and
the second wave packet is for the second skip.  There is very little
difference between umbra, penumbra, and quiet Sun for the first skip,
which is expected because of the large depth below the spot for the
first-skip rays.  There are sizeable differences
between the quiet Sun times and the
spot times for the second skip, which might be expected.  Differences
are seen in both envelope and phase travel times and the covariance
amplitudes.
}
\label{Figcv}
\end{figure*}


As a first step, covariances were averaged separately over the umbra,
penumbra, and the quiet-sun analysis.  The results for two frequency
bandpasses (centered at $3.1\rm{\,mHz}$ and $4.0\rm{\,mHz}$) are
shown in Fig.~\ref{Figcv}.
The filters are Gaussian with full-width-half-maximum (FWHM) of $1.0\rm{\,mHz}$.
Several features are immediately obvious.  For the first skip, whose
wave packet is near 70 minutes, there is little if no difference between
the spot regions and the quiet Sun.
This is expected because of the large depths of the first skip rays
and is a confirmation of the shallow nature of sunspots.
%
For our range of $\Delta$, the depths of the first skip rays are
in the range $51-104\rm{\,Mm}$.
However, for the second skip the
situation is quite different.  The phase times and envelope times are shorter
for the spot regions than for the quiet Sun for both frequency bandpasses.
In addition, the amplitude of the penumbral covariance is considerably
lower than for the umbra or quiet Sun.
\begin{figure*}
\centering
\includegraphics{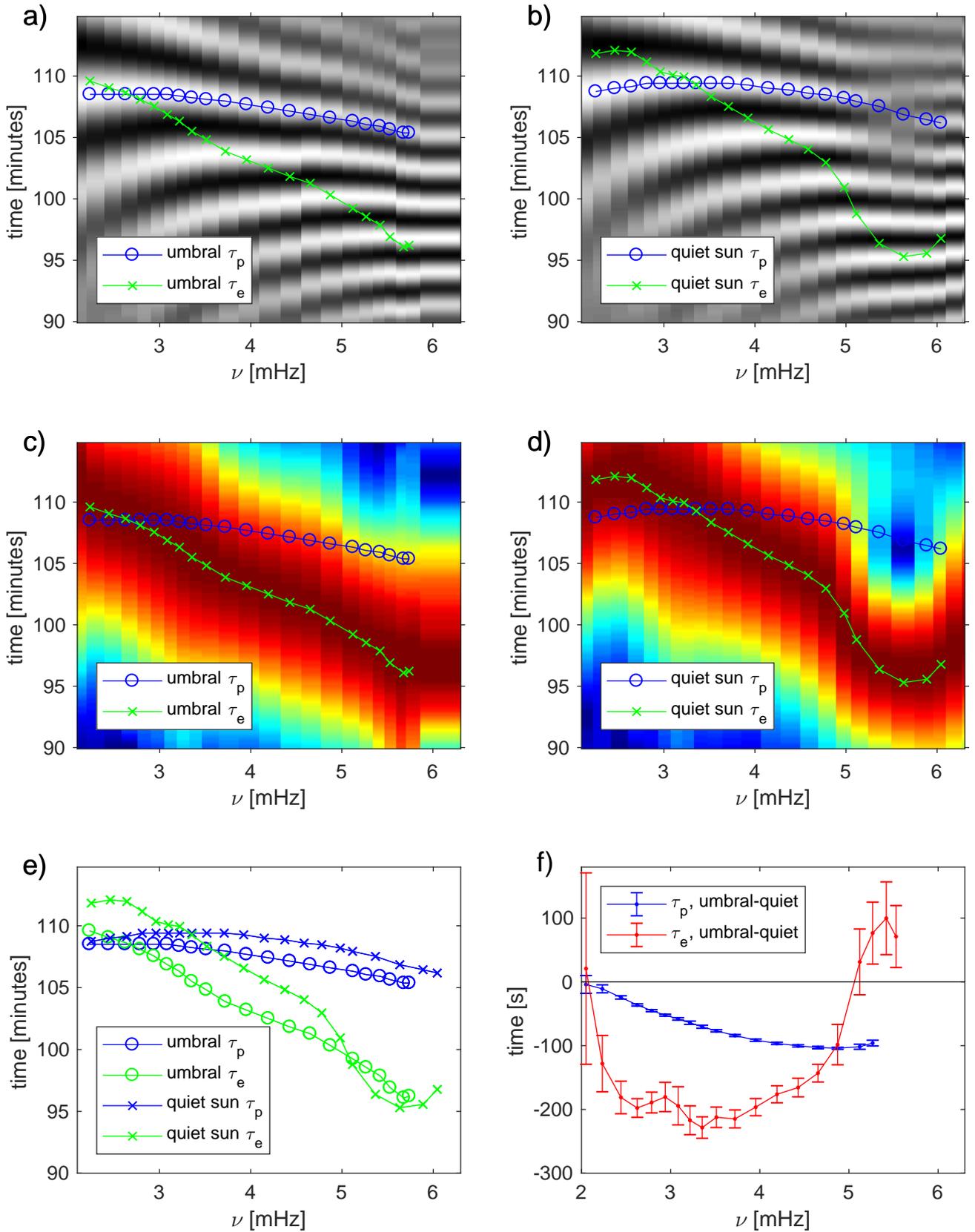}
\caption{two-skip analysis of $\nu$ dependence of the cross covariance for
the umbra (left panels) and quiet Sun (right panels).
Gaussian filters with FWHM=$0.8\rm{\,mHz}$ 
are applied to the cross covariances.
The umbral times are averaged over the full ten days and the quiet sun is
averaged over 9 days.
The grey scale image in the upper left (right) is the $\nu$ resolved
cross covariance
for the umbra (quiet Sun), scaled separately for each $\nu$.
The blue and green curves are the results of the Gabor wavelength fitting for
$\tau_p$ and $\tau_e$.
In the middle row
left (right) plot is shown the envelope of the cross covariance
computed from the analytic signal \citep{Bracewell65} for the umbra (quiet Sun).Overplotted are the same blue and green curves from the top line.
In the lower left, the umbral and quiet Sun $\tau_p$ and $\tau_e$
are shown.  The errors are smaller than the symbols.
In the lower right, the difference travel times (umbral minus
quiet Sun) are shown.
}
\label{Figxc}
\end{figure*}

An important issue is how the waves are reflected below the umbra,
which is related to the depth dependence of the acoustic cutoff 
frequency [$\omega_c$].
This can be studied by
measuring travel times versus the temporal frequency [$\nu$].  For the
quiet sun, this has been done by \citet{Jefferies94} for the envelope
travel times.  For this study,
the average quiet-sun cross-covariance and that for the umbra are
frequency filtered with a Gaussian of FWHM $0.8\rm{\,mHz}$ and subsequently
fit with a Gaussian wavelet \citep{Kosovichev97}
to obtain phase travel times [$\tau_{ph}$]
and travel times for the envelope [$\tau_{env}$].
The distances were averaged over by shifting the correlations for each $\Delta$
relative to the central one.
Fitting results are displayed in Fig.~\ref{Figxc}.

\begin{figure*}
\centering
\includegraphics{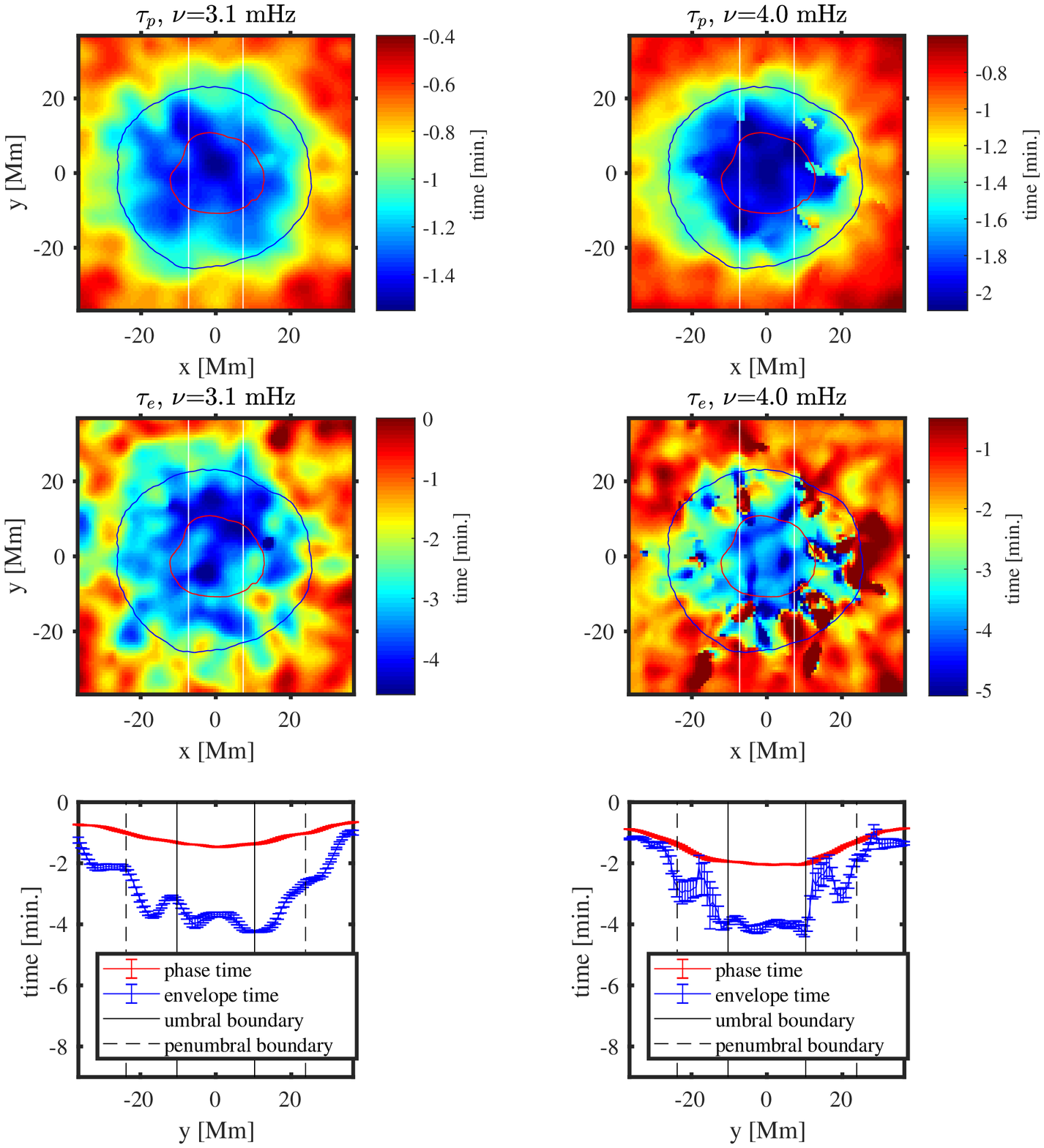}
\caption{Travel-time maps are computed from ten-day averages of cross
covariances.  The cross covariances are $\nu$-filtered with filters
centered at $3.1\rm{\,mHz}$ and $4.0\rm{\,mHz}$ before the travel times are
fitted with
the Gabor wavelets.
Similar 9-day average quiet Sun maps are averaged
over the map and similarly $\nu$-filtered to construct reference
travel times which are subsequently subtracted from those of the sunspot
maps.
The phase (envelope) times $[\tau_p]$ are plotted in the upper (middle)
line for the $3.1\rm{\,mHz}$ in the left column and for the
$4.0\rm{\,mHz}$ in the right column.
Overplotted on the upper four maps are the contours of the umbral-penumbral
boundary (red) and the penumbral-photosphere boundary (blue) as shown in
the earlier figure.
In the lower left and lower right are shown cuts in the north-south
direction averaged over the east-west direction between the pair of
vertical white lines shown overplotted on the maps.  E rror bars are
computed from the scatter of the east-west averages.  Also overplotted
on these cuts are the average umbral-penumbral boundary (heavy black
lines) and the penumbral-photosphere boundary (thin dashed lines).
}
\label{Figmap}
\end{figure*}

It is likely that there is independent information in the phase times
$[\tau_p]$ and in the envelope times $[\tau_e]$ (see the later section 3.).
In the ray theory, $\tau_p$ is obtained by integrating the inverse
phase velocity along the ray.  $\tau_e$ is obtained by integrating the
inverse of the group velocity along the ray.  A theoretical $\tau_p$
(which might be termed the `true phase speed')
would have a unique value while for our observations the phase time
from the cross covariance is only defined within a period.  A way to
resolve (potentially) this nonuniqueness is to go to the high frequencies
above the peak acoustic frequency at $5.2\rm{\,mHz}$.  The pseudomodes at
high frequencies correspond to purely acoustic waves that propagate
outward through the atmosphere.  The `true' phase peak at high $\nu$ should
become constant with $\nu$.  The phase peaks at larger time will slope down
towards this one while the phase peaks at shorter time will slope upwards
towards the true one.  In addition, the envelope times should also
become constant with $\nu$ at high frequencies and be equal to the
true phase times in what is a purely acoustic situation.

In order to test that we are following a single phase peak from low
to high $\nu$, the two-skip cross covariance is shown in Fig.~\ref{Figxc}a
and Fig.~\ref{Figxc}b.
The cross covariance for each frequency filter is shown normalized to
its peak value so that the falloff of amplitude with $\nu$ of several
orders of magnitude is hidden.
For the umbra (Fig.~\ref{Figxc}a), there is no ambiguity in following the
phase peak.  The phase peak that is near the envelope peak at $3\rm{\,mHz}$ is
normally the one that is followed.  For the quiet Sun (Fig.~\ref{Figxc}b),
the phase peaks get a little confused near $5.5\rm{\,mHz}$ with an extra
feature appearing.  The phase time differences are not computed
for $5.5\rm{\,mHz}$ and above because of this issue.  For the envelope times,
there is a similar problem.

In Fig.~\ref{Figxc}c and Fig.~\ref{Figxc}d,
the envelope of the umbral and quiet Sun
covariances computed by an analytic signal formalism is shown with the
travel times measured from the Gabor wavelet fitting superimposed.
The envelope times should be located at the peak of
the envelope computed in this way.  It is immediately apparent that
the dip in $\tau_e$ for the quiet Sun (Fig.~\ref{Figxc}d) near
$5.5\rm{\,mHz}$ is not present for the umbra (Fig.~\ref{Figxc}c).  This dip
was observed in three separate ways for the quiet Sun by \citet{Jefferies94}.

The travel times
$\tau_e$ and $\tau_p$ for the umbra and quiet Sun are compared in
Fig.~\ref{Figxc}e.  The $\tau_p$ for the umbra are in general shorter
than for the quiet Sun.  Except near the confusing region of $5.5\rm{\,mHz}$,
this is also true for $\tau_e$.  The current interpretation of these
shorter times is that the waves are reflected at a lower geometrical
level in the umbra implying a shorter path length and hence shorter
times \citep{Lindsey10}.

It may be useful to consider the quiet Sun times as a reference and
to take the difference of umbra minus quiet Sun.  These differences
for $\tau_e$ and $\tau_p$ are shown in Fig.~\ref{Figxc}f.  The
$\tau_p$ are very well determined.  Both $\tau_e$ and $\tau_p$  become
small near $2\rm{\,mHz}$.  Presumably this is because the waves are reflected
below the sunspot and so these waves do not 'see' the sunspot.  This
suggests a way to avoid the effects of solar activity when trying to
measure global properties like meridional circulation.  That would be
to observe at low frequencies.  This seems difficult in time-distance
analysis as the signal becomes noisy.

Spatial maps of the travel times referenced to the quiet Sun are
shown in Fig.~\ref{Figmap}.  The two bandpasses discussed previously,
centered at 3.1 and $4.0\rm{\,mHz}$ were used.  The phase times are much less
noisy than the envelope times, which we would expect.  The phase times
for $4.0\rm{\,mHz}$ are roughly a factor of two larger than those at
$3.1\rm{mHz}$ in
the umbra, in agreement with Fig.~\ref{Figxc}.
It is interesting that the phase and envelope times are still significant
at the edges of the field.  This is not the case for the theoretical
analysis of the next section.  
A major uncertainty about these maps is what is the horizontal
resolution?  It is possible that the edge effects are caused by poor
horizontal resolution.  Or it may be related to the acoustic moat
reported by \citep{Braun98}.
It would be useful to know how far the travel times are detectable
which could be done by extending the maps.  





\section{Ray simulation} \label{raysec}
To better understand the results of our two-skip analysis of solar data, we perform numerical experiments on the model sunspot of \citet{PrzSheCal15aa} using standard magnetohydrodynamic (MHD) ray theory as described by \citet{MorCal08ab} and \citet{NewCal10aa} in plane-parallel geometry for example, founded on the dispersion function
\begin{multline}
\calD=\omega^2 \omega_{\rm c}^2 a_p^2  k_{\rm h}^2 + 
(\omega^2-a^2\kpar^2)\\ 
\times\left[\omega^4-(a^2+c^2)\omega^2 k^2+a^2c^2k^2\kpar^2 + c^2N^2 k_{\rm h}^2
-(\omega^2-a_z^2k^2) \omega_{\rm c}^2\right] .   \label{DF}
\end{multline}
Here $c$ and $a$ are the sound and Alfv\'en speeds respectively, $\omega$ is the circular frequency, $a_z$ is the vertical component of the Alfv\'en velocity, and $a_p$ is the component perpendicular to the plane containing wave vector $\mathbf{k}$ and gravitational acceleration $\g$. The {\bv} frequency $N$ is defined by $N^2=g/H-g^2/c^2$ where $H$ is the density scale height, $ \omega_{\rm c}$ is the acoustic cutoff frequency, and $ k_{\rm h}$ and $\kpar$ are the horizontal and field-aligned components of the wave vector respectively. 

The associated ray equations are
\begin{equation}
\deriv{\x}{\tau} = \pderiv{\calD}{\boldk}, \quad \deriv{\boldk}{\tau} = -\pderiv{\calD}{\x}, \quad \deriv{t}{\tau} = -\pderiv{\calD}{\omega}, \quad \deriv{S}{\tau}=\boldk\thth\vdot\deriv{\x}{\tau}, \label{ray}
\end{equation}
where $\x=(x,y,z)$ is position, $S$ is phase, $t$ is time, and $\tau$ parametrizes a ray.

The sunspot model is magnetohydrostatic and axisymmetric, based on the method of \citet{KhoCol08aa}, and has been tuned to be both spectropolarimetrically and helioseismically quite realistic, within the confines of the static axisymmetric assumption. Based on the continuum formation height of 5000{\,\AA} radiation, the umbral centre representing the Wilson Depression is at $z=-600$ km, where the magnetic field strength is 3.09 kG.\footnote{The $z=0$ height represents an estimate of the radius of the solar surface obtained from the quiet Sun background model. However, it is slightly offset from the observed surface caused by minor changes in the synthesized continuum intensities obtained from the sunspot model. Our quiet Sun $\log(\tau_{5000})=1$ surface is actually at about $z=-49$ km. } The model does not contain a ``penumbral shelf'', with the magnetic and thermal features being continuous and smooth. For purposes of interpretation, the umbral radius ($R_\text{umbra}=6.6$ Mm) is characterized by $B_z=1.86$ kG \citep{JurBelSch15aa}, and we have arbitrarily identified the edge of the penumbra with the radius where the continuum formation height drops to $-70$ km ($R_\text{penumbra}=19.7$ Mm). The spot is centred at $x=0$, $y=0$ in a cartesian coordinate system. Curvature of the Sun is neglected, which will have little effect as it is travel time differences produced by near-surface perturbations that we work with, rather than raw travel times.

No attempt has been made to adjust the sunspot model to fit the AR11899 spot analysed in Section \ref{data}. Exact correspondences therefore cannot be expected. Nevertheless, broad correspondences (and contrasts) will prove instructive.

As pointed out forcefully by \citet{SchFle98aa,SchFle03aa}, there is no unique acoustic cutoff frequency $\omega_c$ in general. It depends on which variables are used in expressing the wave equation, and the way in which the eikonal approximation is applied. The two most commonly used formulae are the so-called ``isothermal'' cutoff frequency, $\omega_c=\omega_I=c/2H$, and the form of \citet{DeuGou84aa}, $\omega_c^2 = \omega_{DG}^2=(c^2/4H^2)(1-2H')$. The dimensionless number $H'=\rd H/\rd z$ is negative, roughly $-0.5$, throughout most of the convection zone, so $\omega_{DG}\sim1.4\,\omega_I$ in the interior. They are more comparable in the low atmosphere, and in fact are identical in an isothermal atmosphere where $H'=0$. The isothermal form arises naturally in the derivation of the dispersion relation in the appendix of \citet{NewCal10aa}, but that is because only leading order terms in variations of the background ``slowly varying'' atmosphere are retained, so $H'$ does not appear. We mostly employ $\omega_{DG}$ throughout, taking care to smooth the tabulated atmosphere where necessary to avoid unphysical wild oscillations, as it is certainly more firmly founded for the non-magnetic case. However, some comparisons derived with $\omega_I$ are also presented. The main effect of using $\omega_{DG}$ rather than $\omega_I$ is that rays reflect about 200 km lower near the surface, and hence are potentially less affected by the magnetic field.

\begin{figure*}[htbp]
\begin{center}
\includegraphics[width=0.8\textwidth]{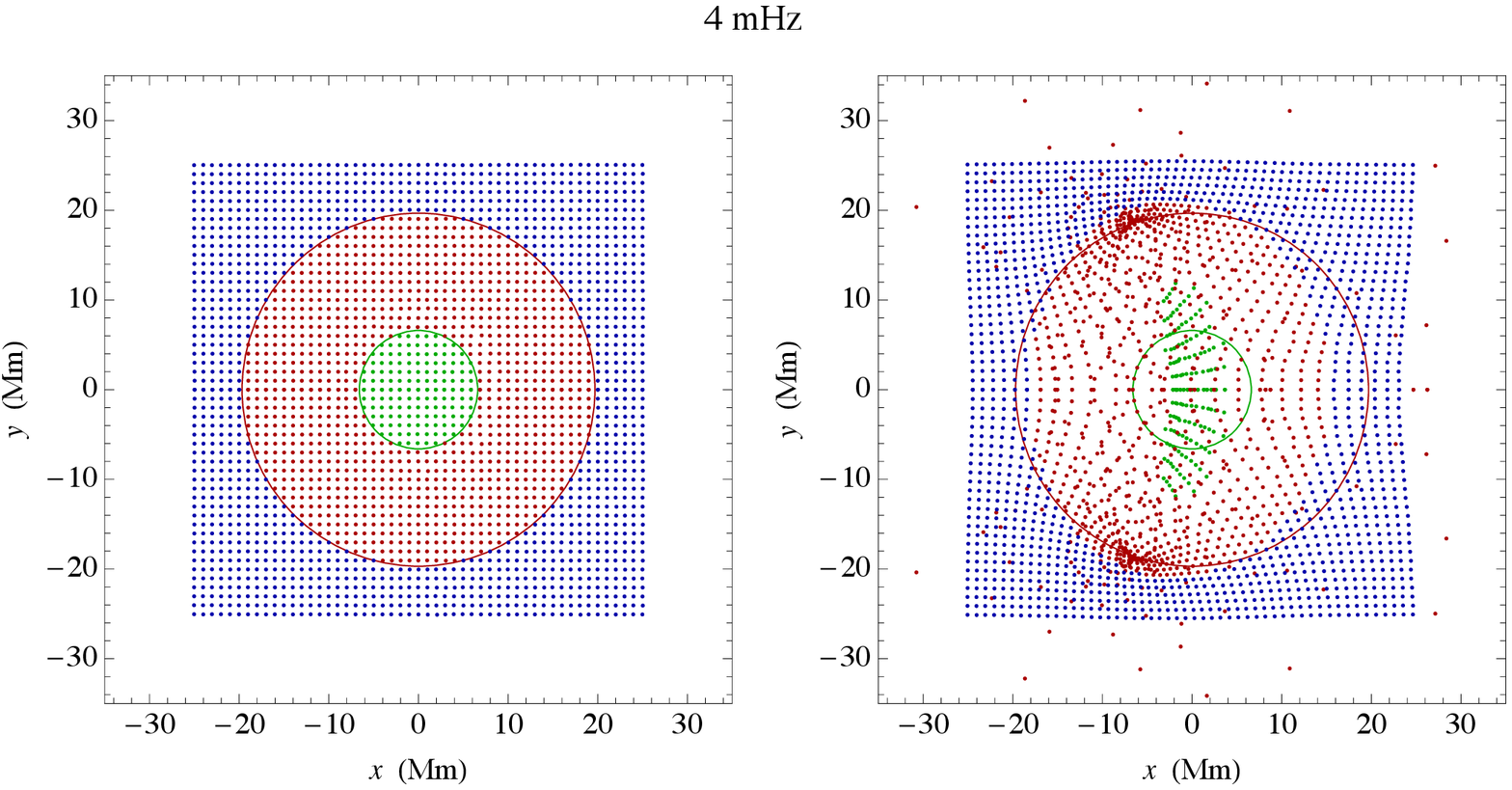}
\includegraphics[width=0.8\textwidth]{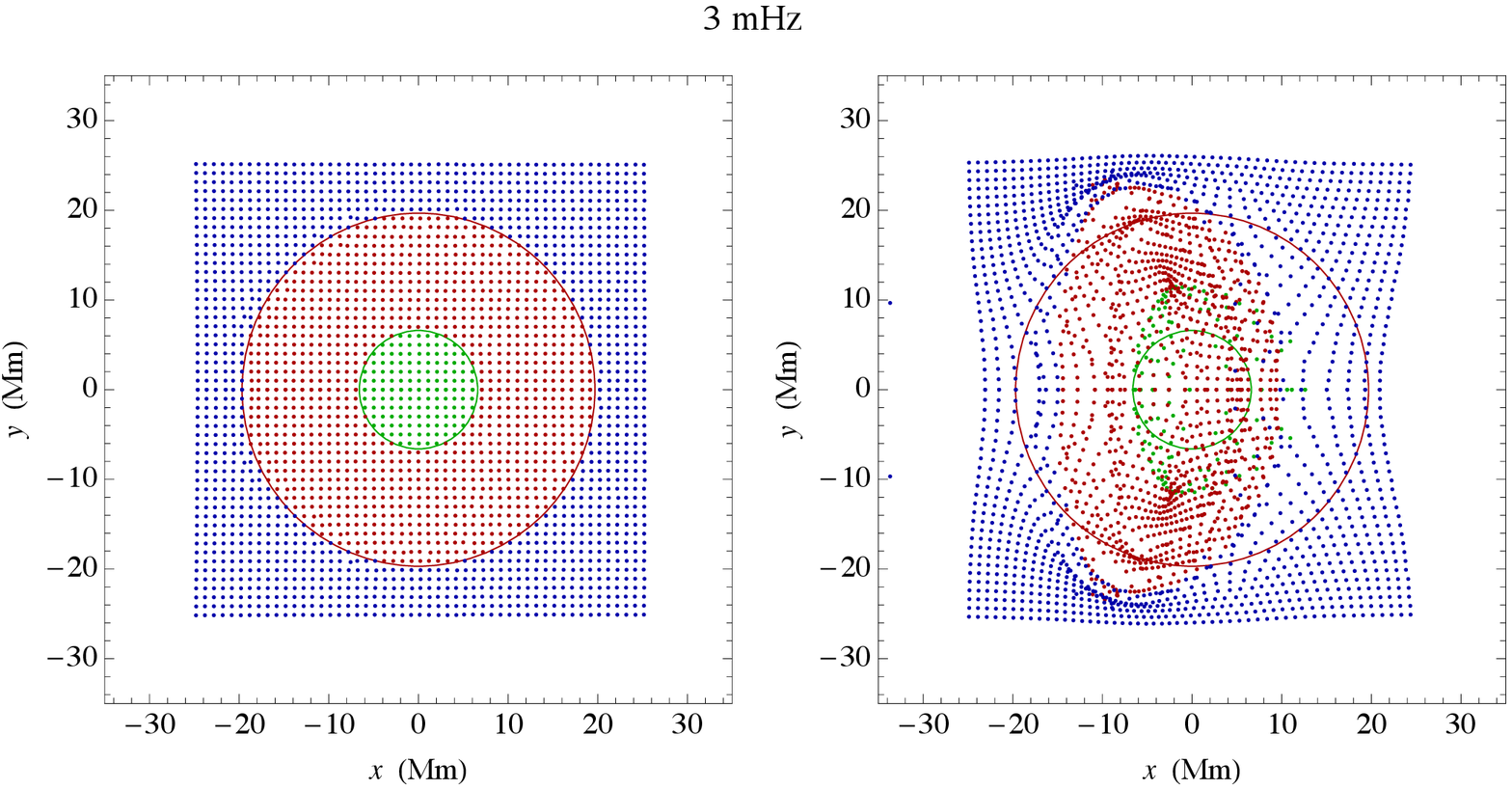}
\caption{Scatter plots of the actual (left panels) and mid-point-inferred (right panels) first-skip turning points for the grid of rays fired from $(-60,0)$ Mm with frequency 4 mHz (top) and 3 mHz (bottom). Points (actually) in the umbra are identified with green colouring, the penumbra with red, and the quiet Sun with blue. The green and red circles are the umbral and penumbral boundaries respectively. These figures use $\omega_c=\omega_{DG}$; scattering with $\omega_c=\omega_I$ is typically substantially increased. }
\label{fig:scatt}
\end{center}
\end{figure*}

Our experiment consists of launching a grid of 3 mHz and 4 mHz rays from their upper turning points at $x=-60$ Mm, $y=0$. These rays are designed to complete their first skip at the integer points of the $(-25\,{\rm Mm},25\,{\rm Mm})\times(-25\,{\rm Mm},25\,{\rm Mm})$ square grid centred on the origin, if the spot is not present. In reality, the spot shifts these points very slightly, as can be seen in the left column of Figure \ref{fig:scatt}. 

On the other hand, the second skip points are displaced, both in direction and skip distance, due to scattering by the spot. Standard practice is to assume the first-skip point is the mid-point between the correlated initial and second-skip points, but the scatter makes this inaccurate. The right column of Figure \ref{fig:scatt} illustrates this by showing where the mid-point-inferred first-skip points would be, though in reality they are as shown in the left column.

The scatter results from the second of Equations (\ref{ray}). The horizontal components of the wave vector $\boldk$ are essentially constant along the ray path, except within typically 100--200 km of the top turning point if it occurs within the sunspot. This is because only in this shallow layer is there a significant horizontal variation in $\calD$, contributed by both the magnetic field and the thermal inhomogeneity. The rays are therefore straight in horizontal projection except for a quite sharp change of direction around the first skip point. Equations (\ref{ray}) are integrated with a high-precision adaptive numerical scheme that follows them accurately through this critical region.

The significant insight from Figure \ref{fig:scatt} is that the sunspot, and in particular the penumbra, substantially scatters the rays. Scatter is much larger if $\omega_I$ is used (not shown) because the rays reach higher into the surface layers and are therefore affected more by the sunspot. Typically, second skip distances and directions are very different from those of the first skip. When the second skip upper turning point is ``observed'' in the quiet Sun, knowing its origin at $(-60,0)$, the standard helioseismic procedure is to infer that the mid-point of these two ends is the central skip point. The figure indicates that this may not be the case, especially for points actually incident in the penumbra, and for the lower frequency. In reality, ``sources'' and ``receivers'' may be oriented arbitrarily with regard to the spot, and these pictures can be azimuthally averaged. Nevertheless, this oriented view is instructive.

Two-skip travel times, both phase and group (envelope) times, are easily recovered from the ray calculations,\footnote{
This is notwithstanding the jump in phase at turning points \citep[][Sec.~5.1]{Bog97aa,TraBriRic14aa}, since only travel time differences are required, and it is assumed that the jump is the same in both cases.} and may be compared with the two-skip times joining the same end-points in quiet Sun (no intervening sunspot). Because of the substantial difference in the physical surrounds of the first skip point in the spot case, and the associated scattering, these times can differ significantly. We define $\delta\tau_\text{ph}$ to be the difference between the two-way two-skip phase travel time through the spot and through the quiet Sun model, and similarly for the group travel time perturbation $\delta\tau_\text{gr}$. Both are typically negative, indicating that the rays pass more quickly through the sunspot than through quiet Sun, despite their often longer ($x$-$y$-projected) path.

\begin{figure*}[htbp]
\begin{center}
\includegraphics[width=.95\textwidth]{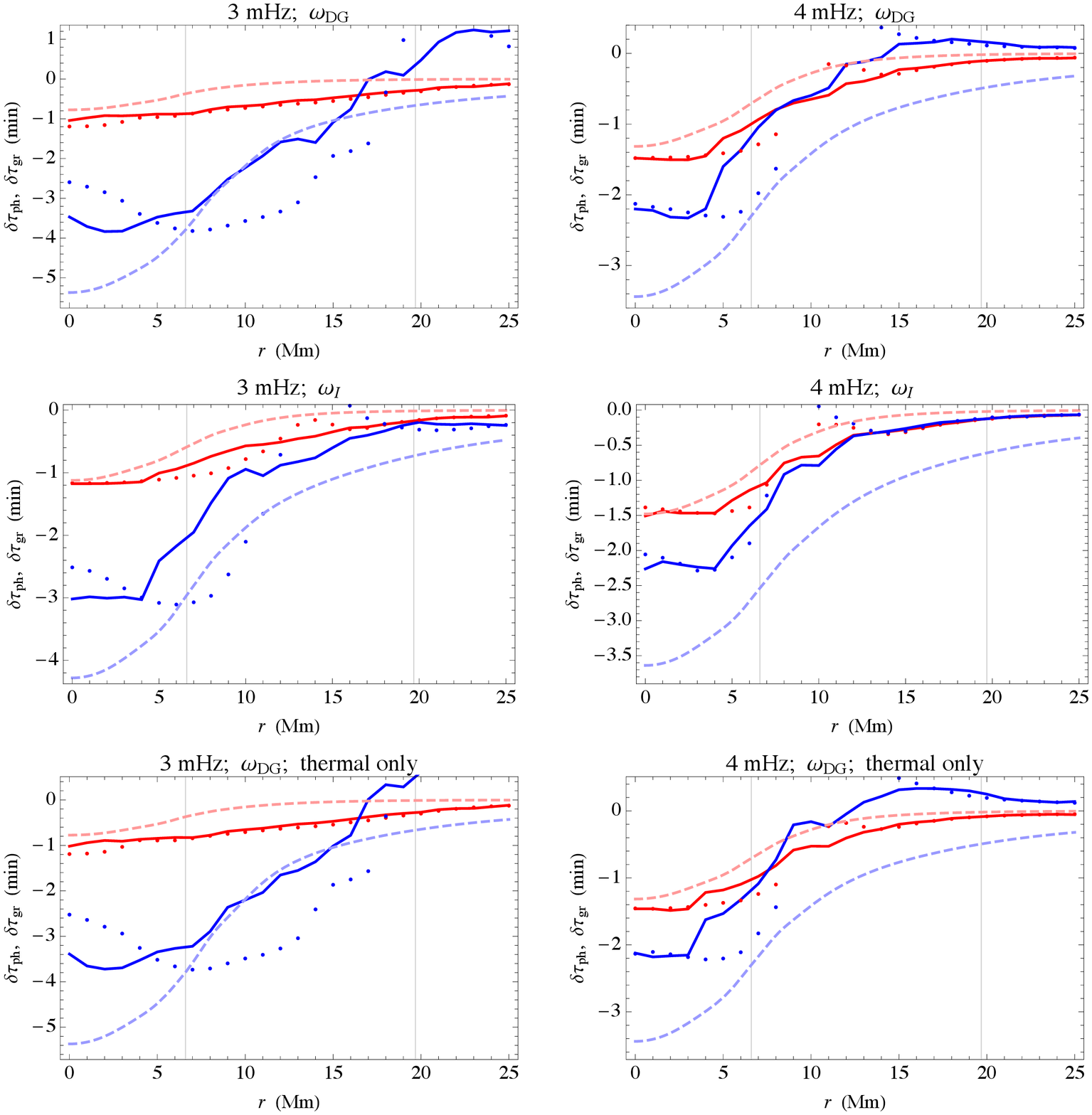}
\caption{Phase (red) and group (envelope, blue) travel time differences, azimuthally averaged, as functions of radius $r$ of the true (dots) or mid-point-inferred (full curves) middle skip points. Left column: 3 mHz; right column: 4 mHz. Top row: full magnetic sunspot model using $\omega_c=\omega_{DG}$; second row: full magnetic sunspot model using $\omega_c=\omega_I$; third row: ``thermal spot'' with the same thermal and density structure, but with magnetic field artificially suppressed. All points were binned to $1\, {\rm Mm^2}$
squares and averaged both by bin and azimuthally. All data presented here has been pre-filtered to remove any rays with second skip distance outside the range $(\frac{5}{7},\frac{7}{5})$ times the first skip distance, or second skip direction more than $20^\circ$ from the first skip direction.  
The fraction of points deleted by this pre-filtering for the six panels is
(0.023, 0.140, 0.122, 0.143, 0.025, 0.129).
The dashed red and blue curves represent respectively the equivalent phase and group speed thermal depths of the Wilson depression; see text for details. The grey vertical lines represent the umbral and penumbral boundaries.}
\label{fig:deltatau}
\end{center}
\end{figure*}


\begin{figure}[htbp]
\begin{center}
\includegraphics[width=0.9\hsize]{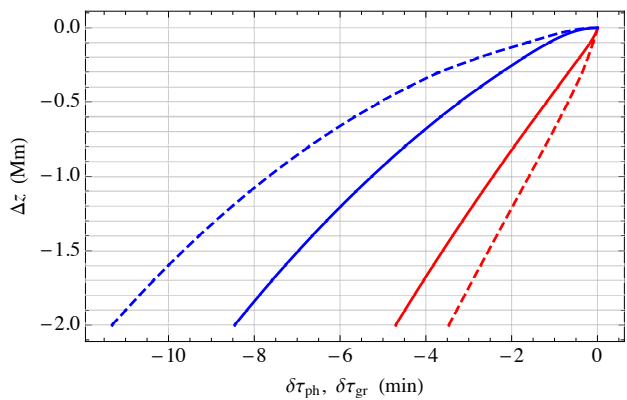}
\caption{Equivalent depths (phase: red; group: blue) for 3 mHz (dashed) and 4 mHz (full).}
\label{fig:depths}
\end{center}
\end{figure}

\begin{figure*}[htbp]
\begin{center}
\includegraphics[width=.9\textwidth]{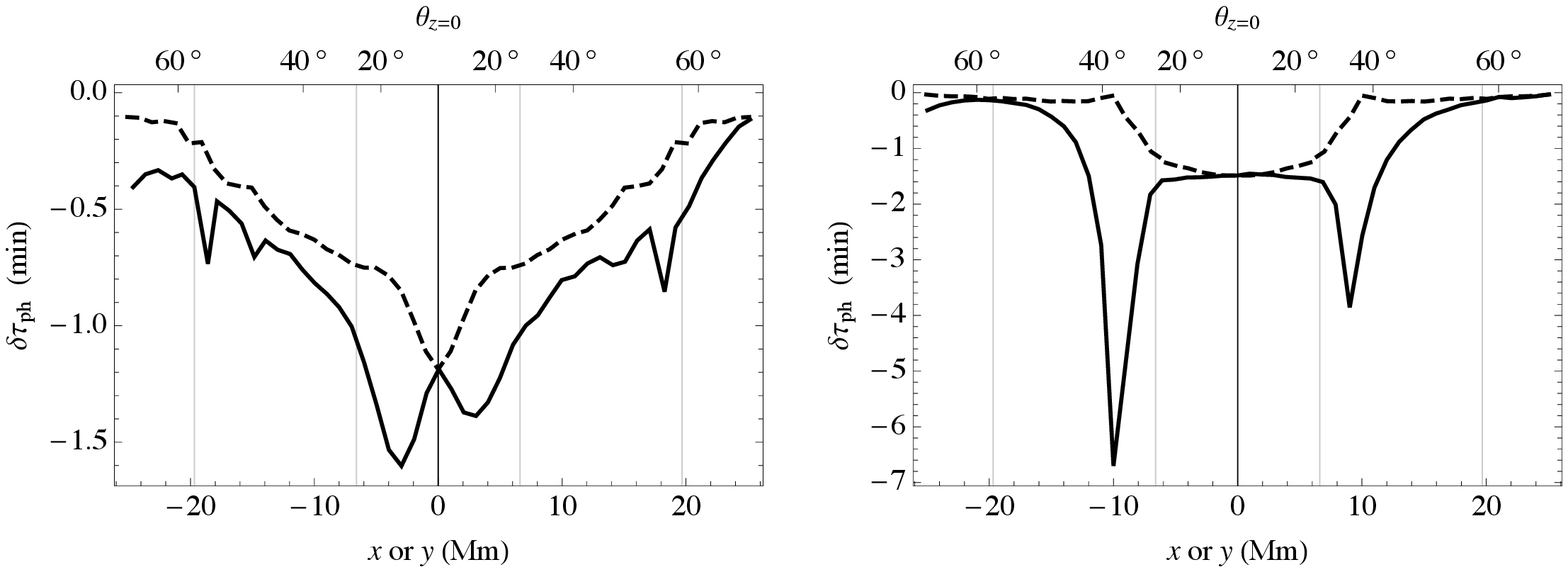}
\caption{Phase travel time perturbations (with $\omega_\text{DG}$) for rays from $(-60,0)$ with first skip turning point lying along the $x$-axis (full curves) and along the $y$-axis (dashed curves). The magnetic field inclination at $z=0$ is indicated on the top axis. Left: 3 mHz; right: 4 mHz.}
\label{fig:phxy}
\end{center}
\end{figure*}

Figure \ref{fig:deltatau} summarizes the timing results. The most prominent points to note are:
\begin{enumerate}
\item Umbral phase travel time perturbations are significantly smaller in magnitude than group travel time perturbations (both are negative).
\item Mid-point-inferred and true centre point travel time perturbations differ substantially in the penumbra, particularly at 3 mHz. This is to be expected given the large degree of penumbral scattering, despite the filter applied to our rays to restrict first and second skip distance contrast to $(\frac{5}{7},\frac{7}{5})$ and direction change to $20^\circ$.
\item The filtering leaves some radii in the penumbra bereft of points, illustrated by gaps in the points representing ``true central point'' travel times. Relaxing the filtering criterion of course fills these gaps, but at the expense of ``true'' and ``mid-point-inferred'' first skip points differing by wider margins.

\item The measured phase times match quite well those predicted by 
the equivalent phase speed depth,
especially in the umbra, where results are more reliable.
 
The group travel time perturbations are consistently smaller than predicted by the equivalent group speed depth.
\item The difference between results obtained with $\omega_{DG}$ and $\omega_I$ at 3 and 4 mHz is quite moderate.
\item There is little substantive difference between results with and without the magnetic field, indicating that the sunspot's thermal structure is primarily responsible for travel time shifts at these frequencies.
\end{enumerate}

The concept of ``the equivalent phase and group speed thermal depths of the Wilson depression'' is a simple though inexact device for converting between Wilson depression depth and travel time perturbations. Given that a ray passes through the surface layers of a sunspot very much faster than through the equivalent depths of quiet Sun \citep[see figs.~3 and 4 of][]{Cal07aa}, the two-way time difference between the magnetic and quiet cases is, to a first approximation, dominated by the quiet Sun travel time: $\delta\tau=-2\int_{z_\text{tp}+\Delta z}^{z_\text{tp}} \rd z/V$, where $V$ is either the vertical phase or group speed, $z_\text{tp}$ is the upper turning point in quiet Sun, and $\Delta z<0$ is the ``Wilson depression'' by which the atmosphere has been lowered in the spot. This correspondence is plotted in Figure~\ref{fig:depths}.

Despite ray travel times being quite insensitive to magnetic field at these frequencies, they are strongly sensitive to direction through inhomogeneities in the background thermal structure, especially at 4 mHz. Figure \ref{fig:phxy} shows phase travel time perturbations along the $x$ and $y$ axes through the spot centre in the magnetic case, with rays launched from $(-60,0)$. The curves hardly differ from the thermal case, indicating that the effect is not directly magnetic. It is instead a consequence of the nature of the scattering on each axis. On the $x$-axis, by symmetry, the only scattering is in second skip distance. Increasing skip distance from the spot centre, the total timing of the now one-short/one-long (or vice versa) two-skip path relative to quiet Sun (symmetric) two-skip times reduces significantly out to about 2 Mm at 3 mHz and 10 Mm at 4 mHz, and then starts to increase as the scattering weakens. On the other hand, along the $y$-axis, the rays largely scatter laterally, thereby reducing the length (and timing) of the required equivalent quiet Sun path, and so the scattered rays' travel time deficits rapidly diminish.


 The ray calculations presented here do not use the ``generalized ray theory'' of \citet{SchCal06aa}, and so do not allow for mode transmission (fast-to-slow; i.e., acoustic-to-magnetic) at the Alfv\'en-acoustic equipartition level. As the 4 mHz rays (for $\omega_c=\omega_{DG}$) barely penetrate the $a=c$ equipartition surface where mode conversion and/or transmission occurs, and 3 mHz rays do not reach it at all, this is unlikely to be of importance in the present context. (With $\omega_c=\omega_I$, some rays reach as high at $a^2/c^2=7$ at 4 mHz.) The effect is much enhanced above 5 mHz, where significant processes involving the atmospheric fast wave are believed to be of importance for both atmospheric waves and interior seismology \citep{CalMor13aa,MorCalPrz15aa,RijMorPrz15aa}. 
 
\begin{figure}[htbp]
\begin{center}
\includegraphics[width=.4\textwidth]{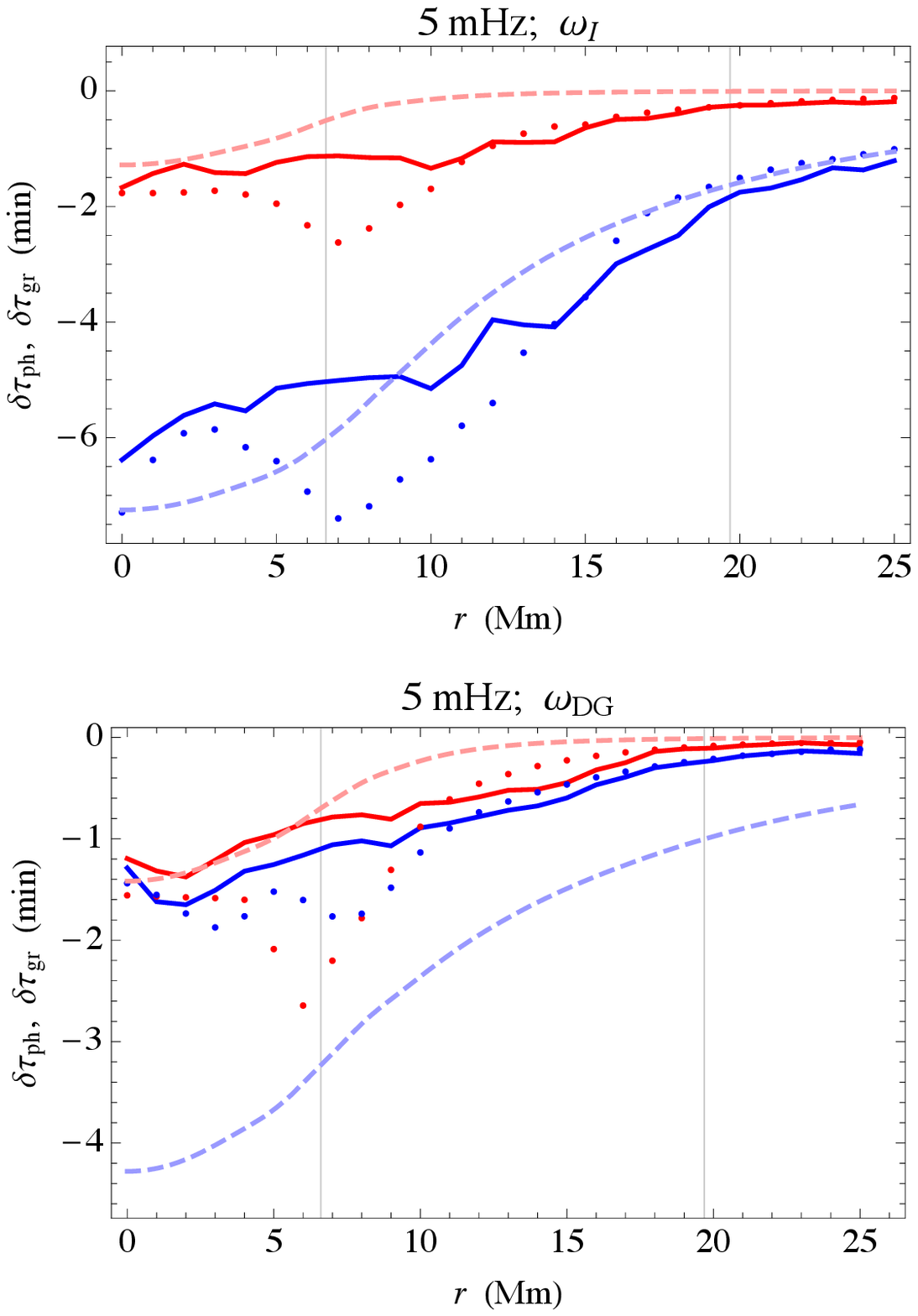}
\caption{Phase and group travel time perturbations, as in Fig.~\ref{fig:deltatau}, but for 5 mHz waves with isothermal (upper) and DG (lower) forms of the acoustic cutoff frequency. Due to increased scatter at this higher frequency, the ray filtering has been relaxed to allow rays with ratio of second to first skip distance in the
range $(\frac{1}{3},3)$ and skip direction change up to $20^\circ$.}
\label{fig:5mHz}
\end{center}
\end{figure}

 Higher frequencies also introduce more uncertainty related to the ``true'' formula for the acoustic cutoff frequency (if such exists). Figure \ref{fig:5mHz} dramatically illustrates the difficulty. Phase and group travel time perturbations are plotted at 5 mHz for each of $\omega_c=\omega_I$ and $\omega_\text{DG}$. The group travel times in particular differ hugely, presumably because at this frequency the rays reach higher in the atmosphere to where the acoustic cutoff formulae differ substantially.  At this stage we do not have a good a priori reason for choosing any of the many alternatives for $\omega_c$, but it is very interesting to note that the $\omega_\text{DG}$ case produces almost identical group and phase travel time differences in the umbra, in accord with observations (Fig.~\ref{Figxc}f). 
 
 A further complication is that the period of a 5 mHz wave is 200 seconds, so a travel time discrepancy of around 400 s (in the top panel) could conceivably have been folded over once or twice observationally. Perturbations that decrease continuously towards zero as $r$ increases (as in Figure \ref{fig:5mHz} ) presumably do not suffer this ambiguity.
 
 
\section{Discussion}

In this paper, measurements of travel times for waves reflecting on the
bottom side of an active region are made and compared with theoretical
calculations of travel times through a sunspot model.  Using the second skip
eliminates the need to use Doppler measurements in the magnetically modified
atmosphere of the active region as was done with center-to-annulus 
distance methods.  
The Fourier-Hankel method \citep{Braun87} and in the more recent method of
correlating the individual location signals with the average over a line
\citep{Cameron08} also do not use the Doppler signal in the sunspot.

The frequency dependence of travel times averaged over the umbra was measured 
and modeled.  The difference of the travel times from the quiet Sun is quite
large.  The envelope time difference reaches a minimum near 3 mHz of -200 s and
is relatively constant in the range 2.5-4.5 mHz.  The phase time difference is
zero near 2 mHz and increases (in magnitude) to -100 s near 5 mHz.  The zero
of the phase time near 2 mHz relative to the quiet Sun suggests that the umbra
is fairly shallow and that the frequency-dependent reflection is below where 
the umbra has an effect.  It would be useful to be able to extend the frequency
range.  For frequencies below 2 mHz, it might be possible to get below 
the sunspot.  At high frequencies it would be useful to have smaller
wavelengths.
However at high frequencies, the waves do not 
reflect and so it is not possible to use the second skip.  At low frequencies, the
background increases as does the horizontal wavelength making useful 
observations difficult. 

One question is how much the travel time signal is reduced at the
center of umbra by the finite wavelengths of the waves used in the
analysis.  The large distances used, $\Delta=6-12\deg$, correspond
to sizable wavelengths at our mapping frequencies of 3.1 and 4.0 mHz.
A simple estimate of the resolution yields an approximate Gaussian
horizontal smoothing of $\sigma=5.6$ Mm (3.1 mHz) and 
$\sigma=4.3$ Mm (4.0 mHz).  If the signal were only due to a constant
Wilson depression over the 11.9 Mm radius umbra, a reduction of the
signal at umbra center of 12\% (3.1 mHz) and 3\% (4.0 mHz) would be expected
from convolving the pillbox-shaped signal with the Gaussian.
Noting the relative flatness of the signal at 4.0 mHz across the
umbra (Fig.~\ref{Figmap}), this seems like a reasonable model for
the umbral Wilson depression and the smoothing.

Additional work is required to obtain more quantitative answers about
the Wilson depression.  Linear simulations of waves traveling through
a sunspot model need to be carried out.  For the largest distance used here,
$\Delta=24\deg$, the depth required of such a model is at least 100 Mm,
somewhat more than has been done to date. 
In the interim, the much cheaper ray calculations of Section \ref{raysec} offer valuable insights, irrespective of the differences between the sunspot model and the real spot.

Comparison of Figs.~\ref{Figmap} and \ref{fig:deltatau} reveals a qualitatively good correspondence in both phase and envelope (group) time delays at around 3 mHz, using the DG acoustic cutoff formula. At 4 mHz the increase in phase time delay is also well-modeled. However, at this higher frequency, the ray calculations underestimate the envelope time delay. At 5 mHz (Fig.~\ref{fig:5mHz}) the difference between phase and envelope delay almost vanishes, both for the real spot and in the ray calculation. It is unclear whether the underestimate in delay at 4 mHz reflects the difference between the model and true spot, or represents a weakness of the ray modeling with this acoustic cutoff formula. Figure \ref{Figxc}f suggests that the envelope delay is maximal around 3 mHz and vanishes around 5 mHz, and that a fairly minor change in the sunspot structure may produce a delay at 4 mHz consistent with Fig.~\ref{fig:deltatau}.

The ray calculations make two striking predictions. The first is that the thermal rather than magnetic structure of the spot is primarily responsible for the two-skip travel time delays. This is testable within linear wave simulations since the magnetic field may simply be turned off with the thermal structure left unchanged 
\citep{Cally09,Lindsey10,Felipe16,Felipe17}.

The second striking insight is the extent to which the rays are scattered both longitudinally (change in $\ell$) and laterally (change in direction). This is harder to test in wave simulations, since the analysis would essentially mimic that used here for the real data. However, with the benefit of better signal-to-noise ratio in simulations, the first-skip point could also be examined, which would allow the hypothesis to be tested. Pending that confirmation, the spatial mapping of the true sunspot assuming the first skip is at the midpoint between the correlated external points must be regarded as suspect. Nevertheless, the true and the midpoint-inferred $\delta\tau$ displayed in Fig.~\ref{fig:deltatau} differ more in detail than in substance.


Our initial success in obtaining observable two-skip phase and envelope travel time differences from AR11899, and relating them to Wilson depression depth via simple ray calculations, suggests that the next important step is to perform large-scale wave simulations with sunspot models of varying depression depth in order to calibrate the correspondence between models and real sunspots. Once this is done, we will have a practically useful new tool for probing spot structure.

\begin{acknowledgements}
The authors would like to thank Aaron Birch and Robert Cameron for useful
discussions.  The HMI data are courtesy of NASA/SDO and the HMI science team.
The data were processed at the German Data Center for SDO, funded
by the German Aerospace Center (DLR).
This work was supported by computational resources provided by the 
Australian Government through the Pawsey Supercomputing Centre under the
National Computational Merit Allocation Scheme, as well as using the gSTAR
national facility at Swinburne University of Technology.  gSTAR is funded by
Swinburne and the Australian Government's Education Investment Fund.
\end{acknowledgements}


\bibliographystyle{aa} 
\bibliography{fred,duvall}

\end{document}